\documentclass[aps,prb,twocolumn,showpacs,floatfix]{revtex4}

\usepackage{amsmath}
\usepackage{amssymb}
\usepackage{bm}
\usepackage{graphicx}

\begin{document}

\title{
Monitoring the localization-delocalization transition within a 1D model with
non-random long-range interaction
}

\author{A.\ V.\ Malyshev}
\affiliation{Departamento de F\'{\i}sica Aplicada, Universidad de Salamanca,
E-37071 Salamanca, Spain}
\thanks{On leave from Ioffe Physiko-Technical Institute,
26 Politechnicheskaya str., 194021 Saint-Petersburg, Russia}
\author{V.\ A.\ Malyshev}
\affiliation{Institute for Theoretical Physics and Materials Science Center,
University of Groningen, Nijenborgh 4, 9747 AG Groningen, The Netherlands}
\thanks{On leave from ``S.I. Vavilov State Optical Institute'',
199034 Saint-Petersburg, Russia.}
\author{F.\ Dom\'{\i}nguez-Adame}
\affiliation{GISC, Departamento de F\'{\i}sica de Materiales, Universidad
Complutense, E-28040 Madrid, Spain}

\date{\today}

\begin{abstract}

We consider a two-parameter one-dimensional Hamiltonian with uncorrelated
diagonal disorder and {\it non-random} long-range inter-site interaction
$J_{mn}=J/|m-n|^{\mu}$. The model is critical at $1<\mu<3/2$ and reveals the
localization-delocalization transition with respect to the disorder
magnitude. To detect the transition we analyze level and wave function
statistics. It is demonstrated also that in the marginal case ($\mu = 3/2$)
all states are localized.

\end{abstract}


\maketitle

Localization-delocalization transition (LDT) in disordered
systems, predicted by Anderson for three dimensions (3D) in
1958,~\cite{Anderson58} (see also Ref. 2) still remains a fascinating problem (see
Refs.~\onlinecite{Lee85,Kramer93,Mirlin00a} for an overview). During the
last two decades, a remarkable progress has been achieved in
understanding the LDT, especially in discovering the nature of the
wave function at transition. This progress became possible thanks
to the fruitful idea of the multifractality of wave functions at
criticality.~\cite{Wegner80,Aoki83,Castellani86,Schreiber91,Janssen94}
This conjecture was then analytically proven for an ensemble of
power-law random banded matrices (PRMB), which revealed the LDT
{\it with respect to the interaction
exponent}~\cite{Mirlin96,Levitov89} (see
Ref.~\onlinecite{Mirlin00a} for an overview). Within the framework
of the latter, it was demonstrated, in particular, that (i) the
distribution function of the inverse participation ratio (IPR) is
scale invariant at transition and (ii) the relative IPR
fluctuation (the ratio standard-deviation/mean) is of the order of
unity at the critical point.~\cite{Evers00,Mirlin00b} This finding
confirmed the conjecture, that was put forward for the first time in
Refs.~\onlinecite{Shapiro86,Cohen88}, that distributions of
relevant physical magnitudes are universal at criticality  (see
also Refs.~\onlinecite{Shklovskii93,Fyodorov95,Prigodin98}). This
invariance is then a powerful tool to monitor the critical point.

In the present paper, we consider a two-parameter tight-binding
Hamiltonian on a regular 1D lattice of size $N$ with {\it
non-random} long-range inter-site interaction:
\begin{equation}
    {\cal H} = \sum_{n=1}^{N}\varepsilon_{n} |n\rangle\langle n| +
    \sum_{m,n=1}^{N}J_{mn}|m\rangle\langle n|\ ,
    \label{hamiltonian}
\end{equation}
where $|n\rangle$ is the ket vector of a state with on-site energy
$\varepsilon_{n}$. These energies are stochastic variables,
uncorrelated for different sites and distributed uniformly around
zero within the interval of width $\Delta$. The hopping integrals
are $J_{mn}=J/|m-n|^{\mu}$, $J_{nn}=0$ with $1 < \mu \leq 3/2$.
For definiteness we set $J>0$, then the LDT {\it with respect to
disorder magnitude} occurs at the upper band edge, provided
$1<\mu<3/2$.~\cite{Rodriguez00,Rodriguez03} The transition is
analogous to that within the standard 3D Anderson model. $\mu=3/2$
represents the marginal case in which all states are expected to
be weakly localized.~\cite{Rodriguez03}

To detect the transition we analyze level and wave function
statistics. We perform a numerical analysis of size and disorder
scaling of the relative fluctuation of both the nearest-level
spacing (LS) and the participation number (PN). The latter is
defined as:
\begin{equation}
    {P_\nu}=\Bigg[\sum_{n=1}^{N} |\psi_{\nu n}|^{4}\Bigg]^{-1}\ ,
    \label{pn}
\end{equation}
where $\psi_{\nu n}$ denotes the $n\,$th component of the $\nu\,$th
normalized eigenstate of the Hamiltonian~(\ref{hamiltonian}).

The relative fluctuation of the nearest-level spacing is  an
invariant parameter at transition, as was conjectured in
Ref.~\onlinecite{Shklovskii93} for the 3D Anderson model and
demonstrated later for a variety of other disordered models (see
e.g. Refs.~\onlinecite{Mirlin00a} and references therein). The
invariance can be used to detect the critical point. We
demonstrate that within the present model, the ratio of the
standard deviation of the PN (SDPN) to its mean value (MPN) is
also an invariant parameter at the critical disorder magnitude
$\Delta_c$. Therefore, the ratio SDPN/MPN can also be used to
detect the transition. To the best of our knowledge, this quantity
has never been used for this purpose.

As the LDT occurs at the top of the band within the considered
two-parameter model, we calculate disorder and size scaling for
uppermost states. Open chains are used in all calculations. We
take advantage of the Lanczos method to calculate the scaling for
large system sizes (up to about $6 \times 10^4$ sites) and two
particular values of the interaction exponent: $\mu = 4/3$ (the
LDT occurs) and $\mu = 3/2$ (the marginal case; no transition is
expected~\cite{Rodriguez03}).
\begin{figure}[ht!]
\includegraphics[width=\columnwidth,clip]{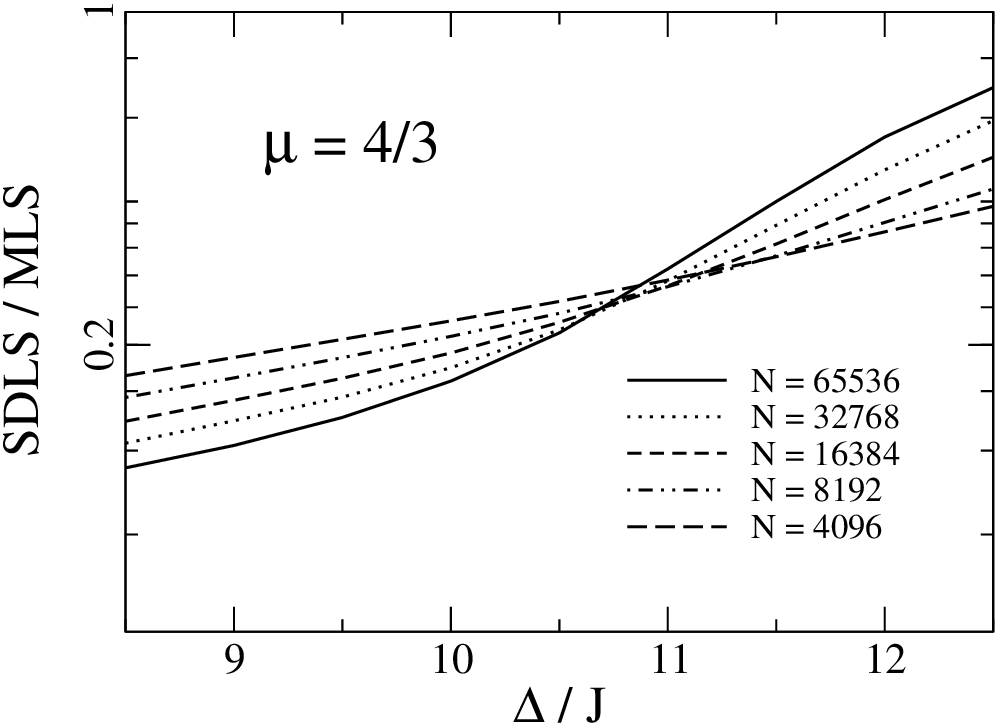}
\caption{
    Disorder scaling of the relative fluctuation of the nearest-level
    spacing (the ratio
    SDLS/MLS) for $\mu=4/3$ in the vicinity of the joint intersection
    point (that is at $\Delta = 10.7\div 11.5 \, J$). The curves are
    calculated for different system sizes $N$ and averaged over more
    than $5\cdot 10^3 \times (65536/N)$ disorder realizations.
    }
    \label{LSTDR_1.3}
\end{figure}

First, we calculate the critical point by means of the level
statistics analysis. In Fig.~\ref{LSTDR_1.3} we plotted the
disorder scaling of the ratio of the standard deviation of the
nearest-level spacing (SDLS) distribution to its mean (MLS) at the
top of the band for $\mu = 4/3$. The figure demonstrates that all
disorder-scaling curves plotted for different system sizes
intersect within a narrow range of $\Delta$, between $10.7J$ and
$11.5J$.

\begin{figure}[ht!]
\includegraphics[width=\columnwidth,clip]{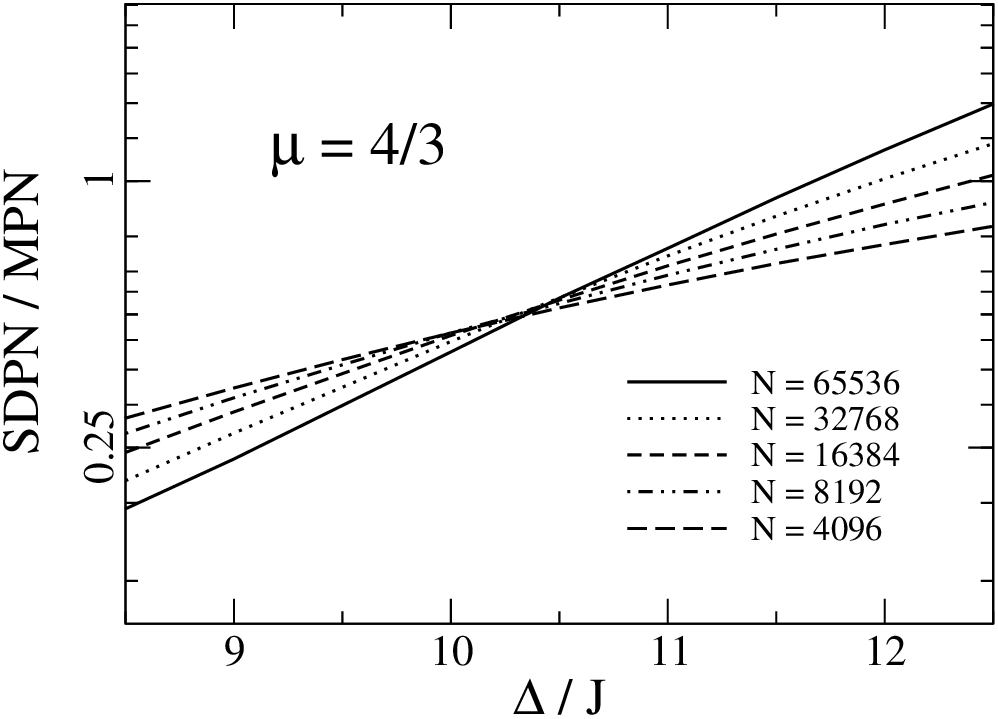}
\caption{
    Disorder scaling of the relative fluctuation of the PN (the ratio
    SDPN/MPN) for $\mu=4/3$ in the vicinity of the joint intersection
    point (that is at $\Delta = 10.0 \div 10.6 \, J$). The curves are
    calculated for different system sizes $N$ and averaged over more
    than $5\cdot 10^3 \times (65536/N)$ disorder realizations.
    }
    \label{STDR_1.3}
\end{figure}

Calculations of the scaling of the relative PN fluctuation confirm
the conjecture that the ratio SDPN/MPN is also a size invariant
parameter at transition: Fig.~\ref{STDR_1.3} shows that all
SDPN/MPN curves plotted versus disorder for different system sizes
intersect in a narrow range of $\Delta$, from $10.0J$ to $10.6J$.
One can deduce from Fig.~\ref{STDR_1.3} that both the MPN and the
SDPN are of the same order of magnitude at the intersection for
any system size, as was shown for other models in
Refs.~\onlinecite{Mirlin96,Evers00,Mirlin00b,Fyodorov95,Prigodin98}.

The regular size dependence of intersection points  in
Figs.~\ref{LSTDR_1.3} and \ref{STDR_1.3} is a finite size effect;
accounting for the latter by means of the finite size scaling
analysis allows for obtaining the value of the critical disorder.
Both figures demonstrate that finite size effects are unusually
strong which, within the present model, results from the
long-range nature of the inter-site interaction. Contrary to the
standard Anderson model, the contribution of the long-range
coupling terms to the spectrum of the Hamiltonian
(\ref{hamiltonian}) converges very slowly as the system size
increases. The latter results in a corresponding increase of the
band width (mostly, the upper band edge, where the LDT takes
place). For an open chain, the upper band edge $E(N)$ size-scales
as follows:
\begin{equation}
    E(N) = E_\infty(\mu) - \frac{C(\mu)}{N^{\mu-1}} + O(N^{-\mu})\ .
    \label{EonN}
\end{equation}
For $\mu=4/3$, $E_\infty(4/3)\approx 7.20J$ and $C(4/3)\approx
8.45J$.  The increase of the band width with the system size leads
to the fact that disorder of the same magnitude is effectively
weaker for larger systems. The latter effect introduces regular
size dependence of the critical disorder that is obtained by
numerical analyses of finite systems. The contribution of other
finite size effects,~\cite{MacKinnon94,Slevin99} such as influence
of boundary regions, are expected to be weaker for large systems
because of very slow convergence of the upper band edge ($\propto
N^{1-\mu}$). Our calculations confirm this conjecture.

The intersection point, $\Delta(N_1,N_2)$, of two disorder scaling
curves plotted for different system sizes  $N_1$ and $N_2$ depends
on the sizes. To account for such dependencies we proceed as
follows. First, set by definition:
\begin{equation}
    \Delta_c(N) = \Delta(N-1,N+1), \quad N_{r} \to \infty\ ,
    \label{DeltacN}
\end{equation}
where $\Delta_c(N)$ is the critical disorder that can be  obtained
by analyses of a finite system of size $N$ ($N_r$ is the number of
disorder realizations over which the averaging is performed).
Second, use the following anzats for the intersection point:
\begin{equation}
    \Delta(N_1,N_2) = w(N_1)\,\Delta_c(N_1)+w(N_2)\,\Delta_c(N_2)\ ,
    \label{DeltaN1N2}
\end{equation}
where the weight function $w(N)$ is to be determined. Bearing in
mind the slow convergence of the band edge ($\propto N^{1-\mu}$),
we use the following anzats for $\Delta_c(N)$:
\begin{equation}
    \Delta_c(N) \approx \Delta_c(\infty)+b\, N^{1-\mu}
    + c\, N^{-\gamma}\ ,
    \quad N\gg 1\ ,
    \label{DeltacApprox}
\end{equation}
where the $b$, $c$, and $\gamma>\mu-1$ are fitting parameters.
Using the anzats (\ref{DeltacApprox}) together with
Eqs.~(\ref{DeltaN1N2}) and (\ref{DeltacApprox}) and expanding
$\Delta(N-1,N+1)$ in series about $N$ (at $N \gg 1$), we find the
weight function: $w(N)=1/2+O(1/N^{\mu+1})$. Further, for any given
pair $N_1 < N_2$ ($N_1,N_2 \gg 1$) there exists the size $N$, such
that $\Delta_c(N)=\Delta(N_1,N_2)$. Making use of the latter
equation together with Eq.~(\ref{DeltacApprox}) and keeping the
leading (non-zero) power of system size in all expansions, we find
the sought $N$:
\begin{equation}
    N = \frac{2^{1/p}\,N_1\,N_2}{(N_1^p+N_2^p)^{1/p}}, \quad p = \mu-1\ .
    \label{N}
\end{equation}
Thus, the intersection point of disorder-scaling curves plotted
for two different system sizes $N_1 < N_2$ ($N_1,N_2 \gg 1$)
yields the critical disorder for an intermediate system size $N$
as defined by Eq.~(\ref{N}).

\begin{figure}[ht!]
\includegraphics[width=\columnwidth,clip]{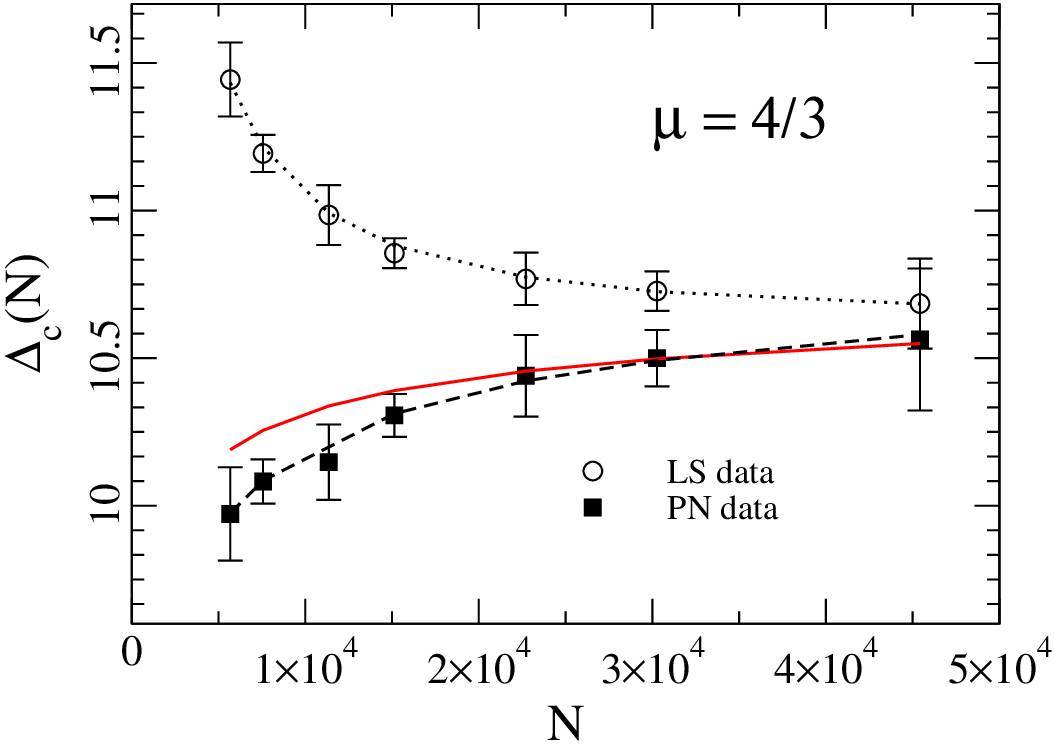}
\caption{
    Critical disorder sizes scaling obtained from the PN data
    ($\blacksquare$) and level statistics data ($\blacktriangle$).
    Dashed lines are best fits of Eq.~(\ref{DeltacApprox}) to
    the LS data ($b=-26.44$, $c=824.16$, $\gamma=0.70$) and
    the PN data ($b=-21.61$, $c=12.73$, $\gamma=2.82$).
    The solid line is the best linear fit of Eq.~(\ref{EonN})
    to the last three PN data points: $\Delta_c(N)=(1.52\pm 0.02)\times E(N)$.
    }
    \label{IntN}
\end{figure}

We further use Eq.~(\ref{N}) and intersection points of the
curves in Fig.~\ref{LSTDR_1.3} (LS data) and Fig.~\ref{STDR_1.3}
(PN data) to obtain $\Delta_c(N)$. Figure~\ref{IntN} shows
$\Delta_c(N)$ together with best nonlinear fits of
Eq.~(\ref{DeltacApprox}) to the whole data sets (dashed lines) and
the best linear fit of $E(N)$ given by Eq.~(\ref{EonN}) to the
three last PN-data points (solid line). The nonlinear fits give
$\Delta_c(\infty)=(10.97 \pm 0.09)J$ for the LS data, and
$\Delta_c(\infty)=(11.19 \pm 0.10)J$ for the whole PN data set,
while the linear fit of $E(N)$ to the tail PN points gives
$\Delta_c(\infty)=(10.91 \pm 0.17)J$. The obtained values of
$\Delta_c(\infty)$ agree well with each other. This confirms our
conjecture that for large system sizes the band edge size
dependence provides the dominant contribution to the finite size
effects. Finally, the critical disorder is determined as
$\Delta_c(\infty) = (11.09 \pm 0.21)J$ for $\mu=4/3$.

It should be noticed that, despite that both methods to detect the
LDT are in good agreement, finite size effects are more
pronounced in the case of analysis of the level statistics. For
these reasons, the proposed method which is based on the studies
of the wave function statistics appears to be advantageous, at
least for the considered model.
\begin{figure}[ht!]
\includegraphics[width=\columnwidth,clip]{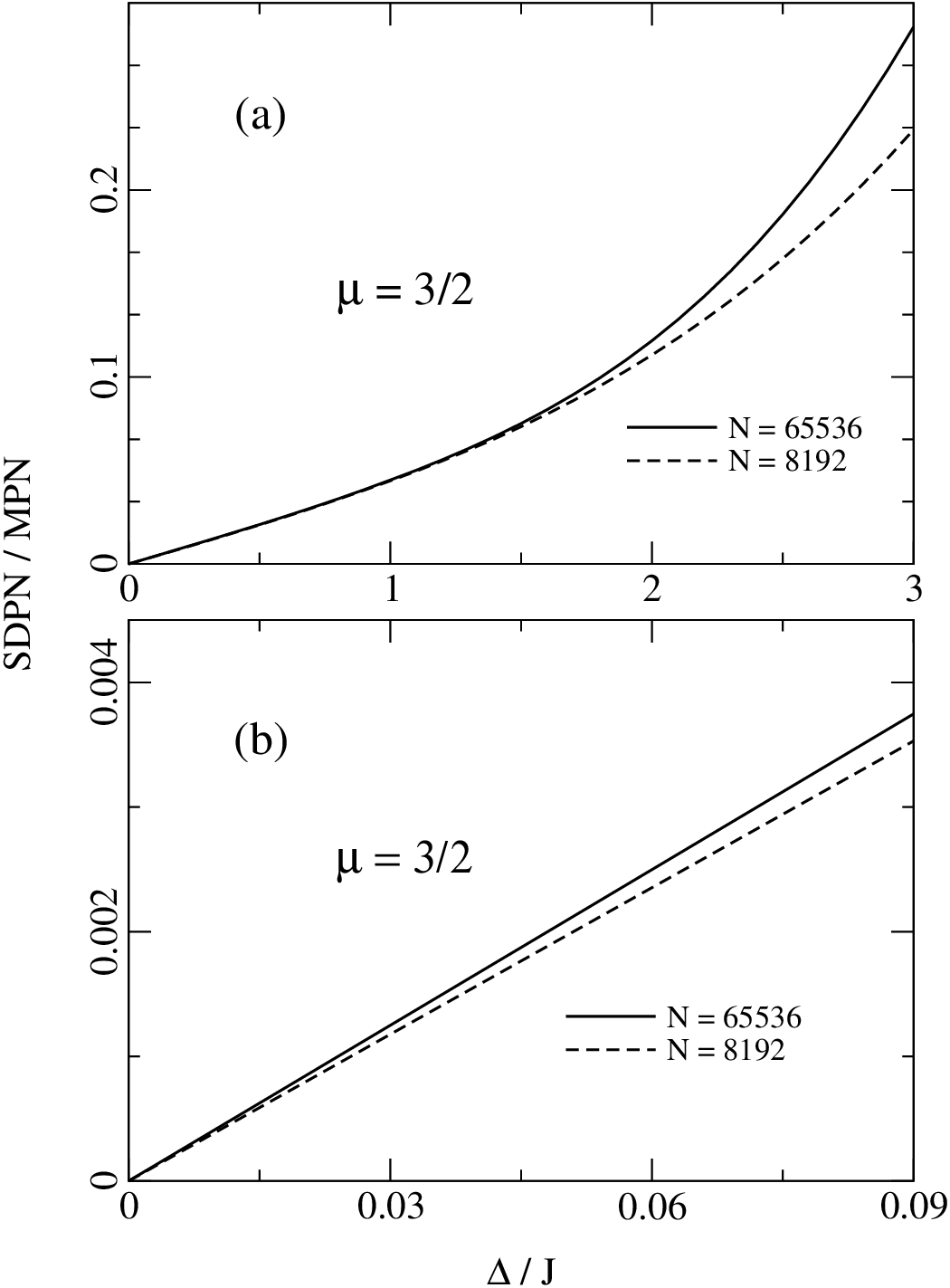}
\caption{
    a) Disorder scaling of the relative PN fluctuation (the ratio SDPN/MPN)
    for $\mu=3/2$ in the vicinity of the joint intersection point at
    $\Delta_c = 0$. The curves are calculated for two different system
    sizes ($65536$ and $8192$) and averaged over more than $5\times10^3$
    and $10^5$ disorder realizations respectively.
    b) A blow up of the crossing at the origin.
    }
    \label{STDR_1.5}
\end{figure}

\begin{figure}[ht!]
\includegraphics[width=\columnwidth,clip]{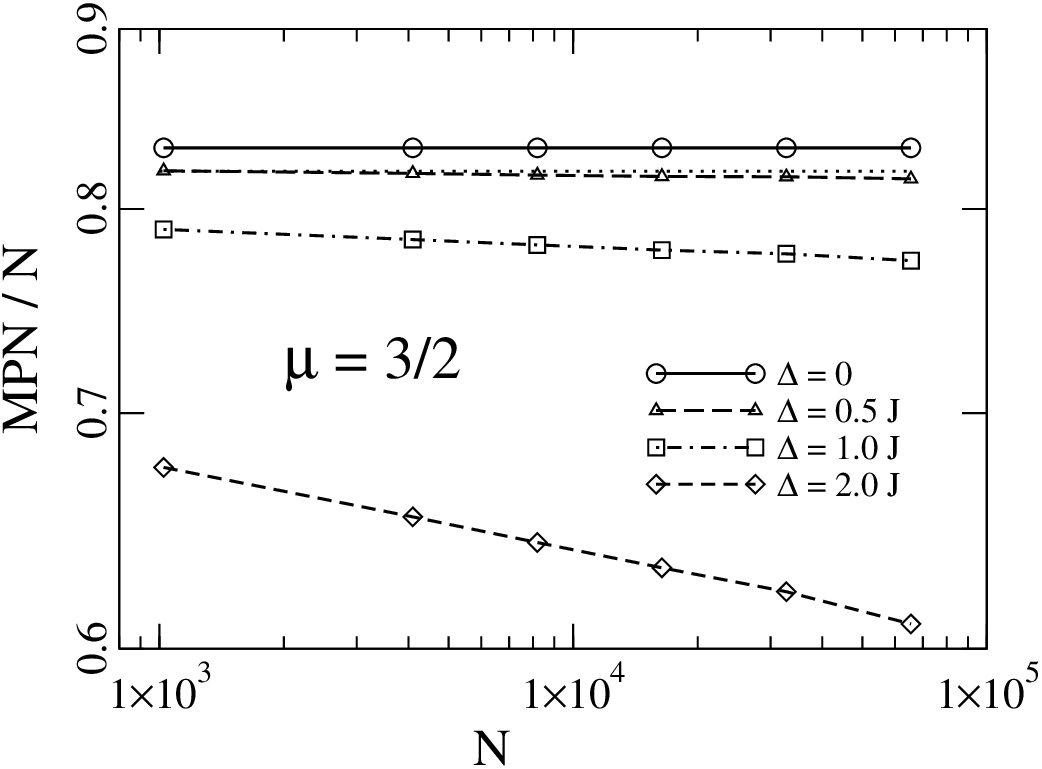}
\caption{
    Size scaling of the ratio MPN/N calculated for $\mu=3/2$ and
    different disorder magnitudes. Thin horizontal dotted line is a
    guide for the eye.
    }
    \label{MPN_N}
\end{figure}

We applied the same technique to analyze the localization
properties  in the marginal case, $\mu=3/2$, where the states are
expected to be localized weakly.~\cite{Rodriguez03}
Figure~\ref{STDR_1.5} shows the SDPN/MPN scaling curves in the
vicinity of the only joint intersection point that appears to be
trivial: $\Delta_c = 0$. Size scaling of the ratio MPN/N (see
Fig.~\ref{MPN_N}) reveals no transition too; all MPN/N
size-scaling curves for non-zero magnitude of disorder decrease
with system size, as they do for a localized (or critical) state.
Thus, no signatures of the LDT can be observed in the marginal
case, indicating that all states are localized.

In summary, we studied numerically the critical properties of the
1D two-parameter tight-binding model with diagonal disorder and
{\it non-random} long-range interaction, $J_{mn} =
J/|{m-n}|^{\mu}$, $J > 0$ and $1 < \mu \le 3/2$. The transition
point was detected by means of the level and wave function
statistics. We used the conjecture on the scale invariance of the
distribution function of the nearest-level spacing and the
participation number at criticality. We find, in particular, that
the critical point for $\mu=4/3$ is $\Delta_c=(11.09 \pm 0.21)J$.
In the marginal case ($\mu=3/2$), that is analogous to the
standard 2D Anderson model,~\cite{Rodriguez03} the only joint
intersection point is $\Delta_c=0$, indicating that all states are
localized for a finite disorder.

We demonstrated that finite size effects are very pronounced
within the considered model. Level statistics appears to be more
affected by these effects as compared to the participation number
statistics. The dominant contribution to finite size effects is
determined by the size dependence of the band width. To obtain the
critial disorder, we use a reformulated finite size scaling
procedure that is corrected for irrelevant size dependencies.

To conclude, we stress that the scale-invariance of the  relative
fluctuation of the participation number at transition is a
consequence of critical wave function fluctuations. We conjecture,
therefore, that the analysis of the relative fluctuation of the
participation number provides a general tool to monitor the LDT.
The proposed method proves to work well for the standard 3D
Anderson model too.~\cite{Schubert03} We believe also that this
property holds at the mobility edge, allowing therefore to monitor
the latter.

\begin{acknowledgments}

The authors thank A.\ Rodr\'{\i}guez, M.\ A.\ Mart\'{\i}n-Delgado and G.\
Sierra for discussions. This work was supported by DGI-MCyT (MAT2003-01533)
and MECyD (SB2001-0146).

\end{acknowledgments}

\end{document}